# Study of longitudinal coherence properties of pseudo thermal light source as a function of source size and temporal coherence

AZEEM AHMAD[1,2,3], TANMOY MAHANTY[1], VISHESH DUBEY[1,2], ANKIT BUTOLA[1], BALPREET SINGH AHLUWALIA[2], DALIP SINGH MEHTA[1,4]

[1]Department of Physics, Indian Institute of Technology Delhi, Hauz Khas, New Delhi 110016, India
[2]Department of Physics and Technology, UiT The Arctic University of Norway, Tromsø 9037, Norway
*Corresponding author: [3]ahmadazeem870@gmail.com, [4]mehtads@physics.iitd.ac.in

**In conventional OCT, broadband light sources are generally utilized to obtain high axial resolution due to their low temporal coherence (TC) length. Purely monochromatic (i.e., high TC length) light sources like laser cannot be implemented to acquire high resolution optically sectioned images of the specimen. Contrary to this, pseudo-thermal light source having high TC and low spatial coherence (SC) property can be employed to achieve high axial resolution comparable to broadband light source. In the present letter, a pseudo-thermal light source is synthesized by passing a purely monochromatic laser beam through a rotating diffuser. The longitudinal coherence (LC) property of the pseudo-thermal light source is studied as a function of source size and TC length. The LC length of the synthesized light source decreased as the source size increased. It is found that LC length of such light source becomes independent of the parent laser's TC length for source size of ≥ 3.3 mm and become almost constant at around ~ 30 μm for both the lasers. Thus any monochromatic laser light source can be utilized to obtain high axial resolution in OCT system irrespective of its TC length. The maximum achievable axial resolution is found to be equal to 650 nm corresponding to 1.2 numerical aperture (NA) objective lens at 632 nm wavelength. The findings elucidate that pseudo-thermal source being monochromatic in nature can improve the performance of existing OCT systems significantly.**

Coherence properties of light sources play a crucial role in various optical techniques such as profilometry, digital holographic microscopy, quantitative phase microscopy (QPM) and optical coherence tomography (OCT) [1-5]. Coherence is broadly classified into two categories: temporal and spatial coherence [6-8]. The spatial coherence (SC) is further divided into two sub-categories: lateral and longitudinal spatial coherence (LSC). The coherence gating generated due to the low temporal coherence (TC) length of broadband light sources have been widely utilized in Full-Field OCT (FF-OCT) systems for non-contact and non-invasive optical sectioning of the biological cells/tissues [1, 5]. However, the main drawback while using broad band light sources in FF-OCT is the requirement of dispersion-compensation mechanism for dispersion correction. Further, the highly absorbing biological samples exhibit inhomogeneous spectral response to different wavelengths contained in broadband light source [1, 6, 9]. These issues overall limit the penetration depth and reduces signal to noise ratio (SNR) of FF- OCT system and further add complexity to the system. To overcome these limitations, purely monochromatic light sources like lasers can be implemented in FF-OCT system. However, high TC length of these light sources reduces the depth sectioning capability of the tomography systems [6]. Further, high TC length degrades the image quality due to the speckle noise, coherent noise and parasitic fringe formation.

On the other end, a pseudo-thermal light source, which have high TC and low SC properties, can be advantages over all commercially available light sources: direct lasers and broadband [6, 10]. The pseudo-thermal (or quasi-thermal) light source is an extended monochromatic light source, which is generated due to the coherent light scattering from an optical rough surface. Previously, quasi-thermal light source has been synthesized by passing the laser light through the rotating diffuser [2, 4, 11] or stationary diffuser followed by a vibrating multiple multimode fiber bundle (MMFB) [12-14]. To date, their longitudinal spatial coherence (LSC) properties have been utilized in the field of profilometry, OCT and QPM [4, 6, 10, 13, 15, 16]. A significant number of works have been reported previously for investigating coherence properties of such light sources [11, 17-20]. These types of light sources do not require any dispersion compensation mechanism for dispersion corrections while imaging biological specimens having strong dispersion or inhomogeneous spectral response [4]. In addition, implementation of this light source alleviates the problems of coherent noise, speckle noise, and parasitic fringe formation in FF-OCT systems [4, 18]

In the present letter, a pseudo thermal light source having high TC and low SC properties is synthesized by passing the laser light through a rotating diffuser. The longitudinal coherence (LC) properties of spatially extended monochromatic light sources synthesized from two different monochromatic laser sources: He-Ne (@ 632 nm) and DPSS (@ 532 nm), is systematically studied. The LC length of such pseudo thermal light sources is measured as function of source size by employing Linnik based FF-OCT system and found to minimum for large source size. Moreover, it is also observed that LC length of pseudo thermal light source does not depend on TC length of the parent laser for source size greater than 3.3 mm. At this source size, numerical aperture (NA) of the objective lens (10X and 0.30 NA in our case) becomes dominant over the TC length of the primary laser and leads to high axial resolution of ∼ 15 micron. The axial resolution of the FF-OCT system is further improved by employing high NA (1.2) objective lens and found to be equal to 650 nm. The spatially extended monochromatic light source compasses the advantages of both pure monochromatic laser and the broadband light source. This opens an opportunity to use such light sources for various OCT applications with various advantages.

In the coherence theory of optical fields, Wiener–Khintchine theorem is used for the determination of TC function [7, 8]. Temporal coherence describes fixed or constant phase relationship, i.e., correlation between light vibrations at two different moments of time. According to this theorem, autocorrelation or temporal coherence function $\Gamma(\Delta t) = \langle E(t)E^*(t-\Delta t)\rangle$ and source power spectral density forms Fourier transform pairs and given by the following relation [6, 8].

$$\Gamma(\Delta t) = \int_{-\infty}^{\infty} S(\nu) \exp(i2\pi\nu\Delta t)\, d\nu \qquad (1)$$

where, $\Gamma(\Delta t)$ is the temporal coherence function, $S(\nu)$ is the source spectral distribution function, and $\Delta t$ is the temporal delay between optical fields $E(t)$ and $E^*(t-\Delta t)$.

The generalized van Cittert–Zernike theorem relates LSC function to the spatial structure (i.e., angular frequency spectrum) of the quasi monochromatic extended light source [6]. Analogous to the Wiener–Khintchine theorem [7], the generalized van-Cittert–Zernike theorem [6, 7] states that LSC function '$\Gamma(\delta z, \Delta t=0)$' and source angular frequency spectrum form Fourier transform pairs. The LSC function is defined as follows:

$$\Gamma(\delta z, \Delta t = 0) = \int_{-\infty}^{\infty} S(k_z) \exp(ik_z \delta z)\, dk_z, \qquad (2)$$

where $\Gamma(\delta z, \Delta t=0)$ is the LSC function and $S(k_z)$ is the angular frequency spectrum of the light source, $\delta z\ (= z_1 - z_2)$ is the separation between spatial points $Q_1(z_1)$ and $Q_2(z_2)$ situated in two different observation planes along the propagation direction of field, and $k_z$ is the longitudinal spatial frequency [8].

The general expression of the LC length ($L_c$) which depends on both the angular frequency and temporal frequency spectrum of the light source, as follows [8]:

$$L_c = \left[\frac{2\sin^2(\theta_z/2)}{\lambda} + \frac{\Delta\lambda}{\lambda^2}\cos^2(\theta_z/2)\right]^{-1} \qquad (3)$$

where, $\theta_z$ is half of the angular spectrum width, $\lambda$ is the central wavelength, and $\Delta\lambda$ is related to the temporal spectrum width of the source.

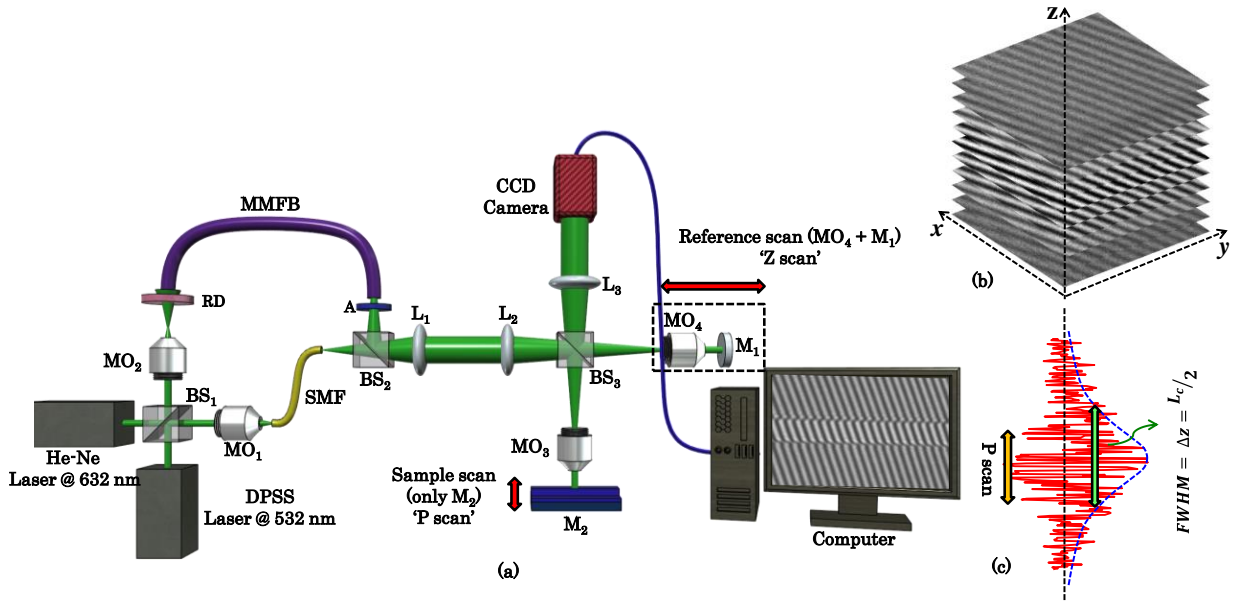

**Fig. 1.** (a) Schematic diagram of the spatial coherence gated FF-OCT system. $MO_{1-4}$: Microscope objectives; $BS_{1-3}$: Beam splitters; $L_{1-3}$: Lenses; RD: Rotating diffuser; MMFB: Multiple multi-mode fiber bundle; SMF: Single mode fiber; A: Variable aperture; M: Mirror and CCD: Charge coupled device. P-scan: vertical scan of mirror $M_2$ for LSC length measurement and Z-scan: horizontal scan of microscope objective $MO_4$ and mirror $M_1$ assembly (enclosed in black dotted box) for the measurement of temporal coherence length of light source. (b) Stack of the recorded interferograms obtained by translating mirror $M_2$ (Fig. 1a) along the vertical direction. (c) Variation of intensity at a particular pixel of interferogram's stack along z- direction for the quantification of FWHM of the visibility curve depicted in blue dotted curve, which provides axial resolution '$\Delta z$' of the system.

The experimental scheme of FF-OCT system is illustrated in Fig. 1a, which is based on the principle of non-common-path Linnik based interference microscopy. A sample mirror is positioned under FF-OCT system for the measurement of LSC length of pseudo thermal light source as a function of aperture size. The laser light beams coming from He-Ne ($l_c$ = 15 cm) and DPSS laser ($l_c$ = 6 mm) are split into two beams using beam splitter $BS_1$. One of the beams goes towards microscope objective $MO_1$ which couples both light beams into a single mode fiber (SMF) to generate spatially filtered temporally high and spatially high coherent light beam. The other one goes towards microscope objective $MO_2$ which illuminates the rotating diffuser (RD) with a diverging beam (spot size at diffuser plane ~ 4 mm). The RD generates temporally varying speckle field and eventually reduces the speckle contrast significantly [21]. The scattered photons are directly coupled into a multiple multi-mode fiber bundle (MMFB) placed at ~ 1 mm distance from the diffuser plane to maximize the number of coupled photons into MMFB. The diameter of MMFB is 5 mm and contains hundreds of fibers (core diameter of each fiber ~ 0.1 mm). The RD followed by MMFB generates uniform illumination, i.e., speckle free field, at the output port of MMFB, which acts as an extended purely monochromatic light source namely pseudo thermal light source. Thus, generates a temporally high and spatially low coherent light source having short LSC length which depends on the spatial extent of the light source.

The output port of MMFB is attached with the input port of interference microscopy system (Fig. 1a). A variable aperture (A) is placed between output port of MMFB and input port of microscope, which controls the SC (lateral and longitudinal both) properties, i.e., LSC length, of the pseudo thermal light source. The light beams coming from SMF and MMBF are recombined using $BS_2$, which directs ~ 50% intensity of both the light beams towards Linnik interferometric system. The combination of lenses $L_1$ and $L_2$ relay the source image (fibers of MMFB) at the back focal plane of the microscope objective $MO_3$ to achieve uniform illumination at the sample plane. The lenses $L_1$ and $L_2$ tightly focus the light beam coming from SMF at the back focal plane of the microscope objective MO3, which sends a nearly collimated beam at the specimen plane. The beam splitter $BS_3$ splits both beams into two; one is directed towards the sample ($M_2$) and the other one towards reference mirror ($M_1$). Both light beams reflected back from $M_2$ and $M_1$ recombine and forms interference pattern at the same beam splitter plane, which is projected at the camera plane with the help of $L_3$. The angle of reference mirror $M_1$ controls the angle between the object and the reference beam, i.e., the fringe width of the interferogram. The reference mirror $M_1$ is kept at a particular angle for which high fringe density without aliasing effect is observed at the change coupled device (CCD) plane. The CCD camera [Lumenera Infinity 2, 1392 × 1024 pixels, pixel size: 4.65×4.65 μm$^2$] is utilized for all interferometric recordings.

For the measurement of LSC length or axial resolution (i.e., LSC/2), a flat mirror $M_2$ (Fig. 1a) as a test sample is placed under the interference microscope and scanned vertically in a constant step (1 and 0.2 μm) from − z to + z (P scan) to sequentially acquire a series of interferograms. The series of interferograms are then stacked along the z direction as presented in Fig. 1b for the measurement of LSC function of the light source. The variation of intensity at a particular pixel of interferogram's stack (Fig. 1b) is plotted as a function of z as depicted in Fig. 1c. The blue dotted curve illustrated in Fig. 1c exhibits the LSC function of pseudo thermal light source. It is observed that the fringe visibility of interferograms reduces as $M_2$ go away from the focal position of microscope objective $MO_3$. The FWHM of the LSC function thus obtained provides information about the axial resolution 'Δz' and subsequently LSC length (=2Δz) of synthesized light source.

A systematic study is done to understand the influence of source size on the LC length of pseudo thermal light source. The size of the source is varied in a step of 0.5 mm from 0.8 mm to 4.8 mm with the help of variable aperture 'A'. Two microscope objective lenses $MO_3$ and $MO_4$ having NA 0.3 (10×) are utilized in Linnik interference microscopy system. The LC length of extended monochromatic light source synthesized from two high coherent lasers: DPSS laser ($l_c$~ 6 mm) and He-Ne laser ($l_c$ ~ 15 cm) is measured as a function of source and compared. Figures 2a – 2h present the normalized LC functions of pseudo thermal light source for different source sizes at two different wavelengths. The LC functions illustrated in green and red color profiles (Figs. 2a – 2h) correspond to 532 nm and 632 nm wavelengths, respectively. A slight asymmetry in the measured LC functions could be due to slight misalignment in the beam path. The FWHM of each LC functions corresponding to 532 nm and 632 nm wavelengths are then calculated to obtain LC length and subsequently axial resolution of the system as given in Table 1.

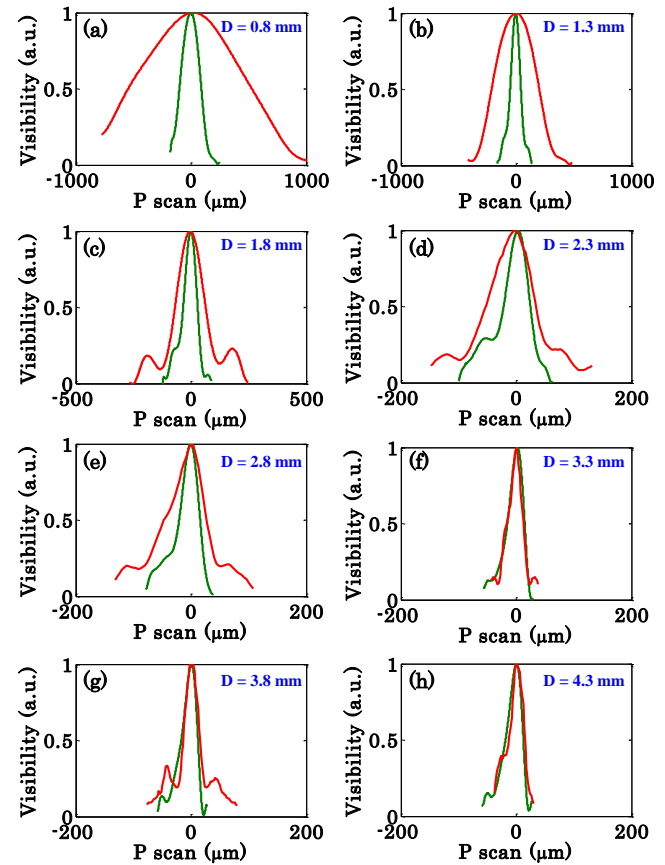

Fig. 2. The LC functions of extended monochromatic light source synthesized from two high coherent lasers: DPSS laser ($l_c$ ~ 6 mm) and He-Ne laser ($l_c$ ~ 15 cm) as a function of source sizes: (a – h) 0.8 – 4.8 mm in a step of 0.5 mm. Green and red color curves represent the LC functions corresponding to 0.3 NA objective lens at 532 nm and 632 nm wavelengths, respectively.

Table 1. Longitudinal coherence (LC) length and axial resolution 'Δz' of the pseudo thermal light source synthesized from DPSS and He-Ne laser as a function of source size.

| Source size (mm) | DPSS Laser @ 532 nm | | He-Ne Laser @ 632 nm | |
|---|---|---|---|---|
| | $L_C$ (μm) | Δz (μm) | $L_C$ (μm) | Δz (μm) |
| 0.8 | 190.4 | 95.2 | 1072 | 536 |
| 1.3 | 94 | 47 | 412 | 206 |
| 1.8 | 62.8 | 31.4 | 132.8 | 66.4 |
| 2.3 | 46.4 | 23.2 | 80 | 40 |
| 2.8 | 36.4 | 18.2 | 64 | 32 |
| 3.3 | 30.2 | 15.1 | 32 | 16 |
| 3.8 | 29.2 | 14.6 | 30 | 15 |
| 4.3 | 31 | 15.5 | 28 | 14 |
| 4.8 | 28.2 | 14.1 | 27 | 13.5 |

It can be clearly seen from Table 1 that LC length decreases as the size of aperture increases. The maximum LC length was obtained for source size of 0.8 mm for both the wavelengths. However, LC length measured at 632 nm wavelength is found to be large compared to 532 nm wavelength. When the large source size ≥ 3.3 mm is used, the LC length becomes independent on the TC length of the parent laser source and approached to a constant value of about ~ 30 μm. This is in synchronous with Eq. 3. For small source size, LC length is decided by both temporal and spatial coherence terms of the source. However, for large aperture size (i.e., wide angular spectrum light source), TC term can be neglected compared to LSC term, thus, LC is purely determined by only LSC term. Fig. 3a and 3b represent the variation of LC length as a function of source size of pseudo thermal light source synthesized from DPPS (@ 532 nm) and He-Ne (@ 632 nm) laser. It is worth noting that LC length first decreases to 30 μm at 3.3 mm source size and then become constant for the source sizes ≥ 3.3 mm. Thus, the LC length is found to be independent on the TC length of parent laser for large aperture sizes.

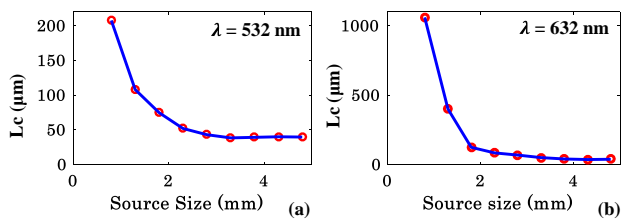

Fig. 3. The LC length of extended monochromatic light source synthesized from two high coherent lasers: (a) DPSS laser @ 532 nm and (b) He-Ne laser @ 632 nm as a function of source size.

The LSC length, i.e., axial resolution, of the system can be further improved with the employment of high NA objective lens as use of high NA widen the angular spectrum light source [6]. To exhibit the improvement in axial resolution, experiments are conducted with water immersion objective lens (1.2 NA) at 632 nm wavelength. The mirror $M_2$ is scanned vertically in a step of 0.2 micron to record a series of interferograms for the measurement of LSC length. Figure 4 presents the visibility curve of the pseudo-thermal light source for 3.3 mm source size. The axial resolution (half of the LC length), i.e., FWHM of the visibility curve, is measured to be equal to 650 nm. The short LC length (~ 1.3 μm) thus achieved may find potential application in high resolution optical sectioning of multilayered biological specimens irrespective of the high TC length of parent laser source.

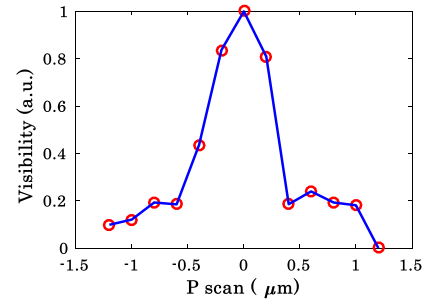

Fig. 4. The LC function of pseudo-thermal light source corresponding to water immersion objective lens (1.2 NA) at 632 nm wavelength and source size of 3.3 mm.

In conclusion, high resolution optical sectioning of the samples is possible with the pseudo thermal, i.e., purely monochromatic (i.e., high TC length) extended light source which is otherwise not possible with the direct laser. This is contrary to the principle of conventional FF-OCT system, which performs high resolution sectioning by utilizing low TC property of broadband light source. The influence of source size on the LC coherence properties of pseudo thermal light source is systematically studied. The axial resolution is measured to be equal to ~ 15 μm for pseudo thermal source size of 3.3 mm and 0.3 NA objective lens. The axial resolution of the system is found to be independent on the TC lengths of the primary laser: DPSS ($l_c$ ~ 6 mm) and He-Ne ($l_c$ ~ 15 cm) lasers at 3.3 mm source size. The axial resolution of the system is further improved by employing 1.2 NA objective lens and measured to be equal to 650 nm at 632 nm wavelength. Thus, use of a sufficiently wide angular spectrum, i.e., low LSC length, synthesized light source enables high resolution sectioning of the specimen irrespective of the TC length of the direct laser. Such light sources can be further implemented for various high axial resolution FF-OCT applications. As already discussed, the use of such light sources are advantageous as it does not require any dispersion compensation and chromatic aberration corrected optics, which are otherwise mandatory in case of broadband light sources [4].

**Funding.** The authors are thankful to Department of Atomic Energy (DAE), Board of Research in Nuclear Sciences (BRNS) for financial grant no. 34/14/07/BRNS. Authors also would like to acknowledge University Grant Commission (UGC) India and Norwegian Centre for International Cooperation in Education, SIU-Norway (Project number INCP- 2014/10024) for joint funding.